\documentclass[journal,10pt]{IEEEtran}
\usepackage{amsmath,amsfonts}
\usepackage{algorithmic}
\usepackage{algorithm}
\usepackage{array}
\usepackage[caption=false,font=normalsize,labelfont=sf,textfont=sf]{subfig}
\usepackage{textcomp}
\usepackage{stfloats}
\usepackage{url}
\usepackage{verbatim}
\usepackage{graphicx}
\usepackage{cite}
\usepackage{amssymb}
\usepackage{epsfig}
\usepackage{rotating}
\usepackage{tabularx}
\usepackage{graphics}  
\usepackage{epstopdf}
\usepackage{multirow}
\usepackage{setspace}
\usepackage{morefloats}
\usepackage{xcolor}
\usepackage{float}
\usepackage[Euler]{upgreek}
\usepackage{mathtools}
\usepackage[mathscr]{eucal}
\usepackage{epsfig}

\begin{document}

\title{\textcolor{black}{Average Age of Information Minimization in Reliable Covert Communication on Time-Varying Channels}}

\author{Shima Salar Hosseini, Paeiz Azmi,~\IEEEmembership{Senior Member,~IEEE,} Nader Mokari,~\IEEEmembership{Senior Member,~IEEE}

\thanks{S. Salar Hosseini, P. Azmi, and N. Mokari are with the Department of Electrical and
	Computer Engineering, Tarbiat Modares University, Tehran, Iran e-mail: ({shima.salarhosseini, pazmi, and
		nader.mokari}@modares.ac.ir)}
\thanks{Manuscript received XXX, XX, 2022; revised XXX, XX, 2022.}}

\markboth{IEEE Transactions on Vehicular Technology,~Vol.~XX, No.~XX, XXX~2022}
{}

\maketitle

\begin{abstract}
	\textcolor{black}{In this letter, we propose reliable covert communications with the aim of minimizing age of information (AoI) in the time-varying channels. 
		We named the time duration that channel state information (CSI) is valid as a new metric, as age of channel variation (AoC).
		To find reliable covert communication in a fresh manner in dynamic environments, this work considers a new constraint that shows a relation between AoI and AoC. With the aid of the proposed constraint, this paper addresses two main challenges of reliable covert communication with the aim of minimizing AoI: 1) users’ packets with desirable size; 2) guaranteeing the negligible probability of Willie’s detection, in time-varying networks. In the simulation results, we compare the performance of the proposed constraint in reliable covert communication with the aim of minimizing AoI with conventional optimization of the requirement of information freshness in covert communications.}

\end{abstract}


\IEEEpeerreviewmaketitle

\section{Introduction}

\IEEEPARstart{I}{n} the sixth generation (6G), secrecy and privacy have key roles such as electronic health records (ERC), financial and identity information transfer, and covering geographical activities in wireless networks \cite{6}. Covert communication is an emerging security technique that aims to protect communications between legitimate transmitter (Alice) and receiver (user), and ensure a low probability of detection by eavesdroppers (Willie) \cite{5}.

In some confidential scenarios, the timeliness of data transmission in status update systems in wireless communications and some applications, such as military, and real-time industrial Internet of things (IIoT) like factory automation, is necessary, because they become outdated \cite{7}. The new performance metric, age of information (AoI) which quantifies the freshness of the received packet at the destination has attracted researchers’ attention. AoI in the covert scheme is the elapsed time since Alice generated the last received update packet at user \cite{8,9,12}.
In \cite{14}, the authors characterize a communication scenario that jointly optimizes the covertness and AoI of short-packets. The objective is to optimize block length to minimize average covert AoI in the additive white Gaussian noise (AWGN) channels.
A covert communications problem with the aim of maximizing the expected detection error probability subject to an average AoI constraint in block Rayleigh fading channels with a full-duplex receiver that transmits artificial noise to confuse the eavesdropper is proposed in \cite{15}.

As mentioned, the goal of covert communications is to prevent detection of communication from Willie's perspective, and non-orthogonal multiple access (NOMA) as a promising technique in 6G, utilizes superposition coding (SC) at Alice will decrease the truthful probability of Willie’s detection \cite{17}. The covert communication in the downlink NOMA system with two users is proposed in \cite{16}. The authors ensure that the strong user receives the covert transmission and the weak user is cover for the strong user. Moreover, a random power transmit scheme is adopted to further confuse Willie.
The covert communication in uplink NOMA is investigated in \cite{19}. Power inversion is controlled by a public user to act as a jammer to prevent detection by the warden, and cover a covert user. 
Furthermore, power domine NOMA (PD-NOMA) allocates different power domain levels to enable multiple users to be served at the same time or frequency band and each subcarrier maps to more than one user, thereby improving spectral efficiency and reducing the average AoI \cite{10,11}. 
The energy-efficient scheduling for AoI minimization in an opportunistic NOMA/OMA downlink broadcast wireless network where users' equipment operates with different quality of service (QoS) requirements is discussed in \cite{18}.

Channel state information (CSI) enables transmissions to be adopted to current channel conditions, which is crucial for achieving reliable communication. CSI is static in coherence time, that is the time duration over which the channel impulse response will not be changed. CSI is therefore outdated after each coherence time duration, hence, \textcolor{black}{we present for the first time the concept of age of channel variation (AoC) which refers to the time duration that CSI is valid.} In order to fulfill covert communications and resource allocation based on channel conditions and user requests, CSI knowledge is necessary. Consequently, Alice practically achieves reliable covert transmission by allocating proper power to get a minimum desirable AoI, depending on the channel variation, and the AoC. It appears that reliable covert communications with the aim of minimizing AoI in the presence of time-varying channels, which is a useful scheme for wireless networks, have been lacking in the literature.


Motivated by the above technical knowledge gap, we propose a reliable covert communication framework with the minimum desirable AoI in time-varying channels. \textcolor{black}{In addition to the advantages of PD-NOMA in 6G that are presented in the literature, PD-NOMA is able to reduce the probability of Willie’s detection by applying superimposing coding at Alice,  and also decreasing the AoI of each requested packet by employing successive interference cancellation (SIC) ordering at each user. We adopt the PD-NOMA technique as transmission technology.} We formulate our objective problem as minimizing AoI in transmitted packets from Alice to all users to jointly specify the optimal Alice power allocation and AoI of each user in order to guarantee the negligible probability of Willie's detection. \textcolor{black}{To achieve reliable covert communication within minimizing AoI for users’ requests with desirable packet sizes, we propose for the first time a new constraint that shows a relation between AoI, and AoC in time varying channels}.
The rest of the subjects are: the transmit power constraint, the covert constraint, and the successful delivery of packets constraint. We exploit an efficient iterative manner to solve the resulting non-linear, and non-convex optimization problem. In our simulation results, we examine the effect of different parameters on our objective function and clarify the effect of the proposed new constraint.

\section{System Model}
\subsection{Considered Scenario and Channel Model}
\begin{figure}[!t]
	\centering
	\includegraphics[width=2.5in]{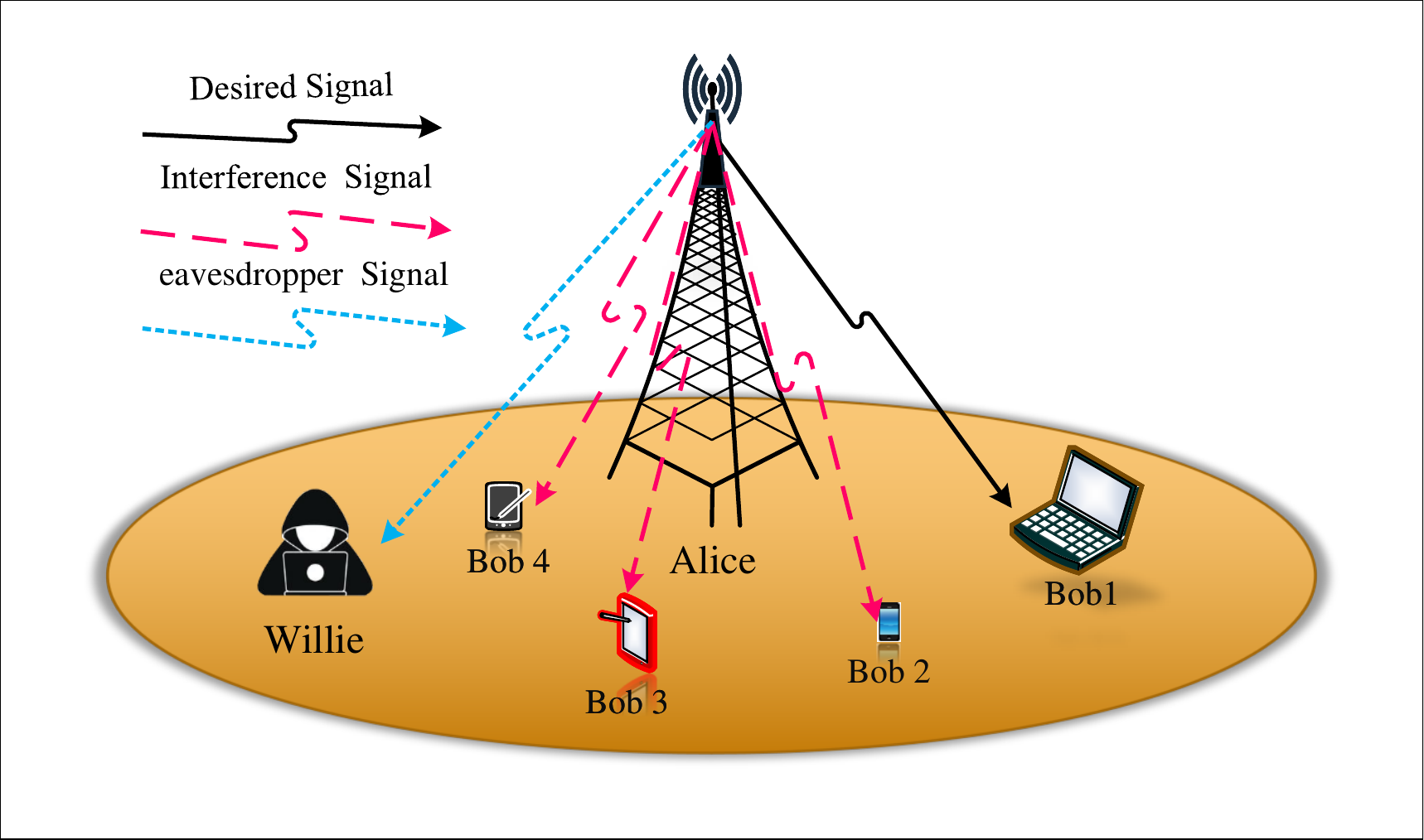}
	\caption{The considered system model.}
	\label{systemmodel}
\end{figure}

We consider a covert PD-NOMA communication scheme where Alice periodically generates packets, and the requested information of users are superimposed and the resulting signal is transmitted towards the users, simultaneously. 
The SIC technique is used to extract the transmitted information. SIC ordering enables users to successfully recover corresponding information such that every user knows the information for which user should be detected and canceled from the received signal.
Alice aims to covertly communicate with all users without being detected by Willie. Willie as an unauthorized user tries to eavesdrop on the transmitted packet over the communication time.
The system model is shown in Fig. \ref{systemmodel}.
We assume that Alice, users, and Willie are equipped with a single antenna. The users set is denoted as $k \in \mathcal{K}=\{1,\dots,K\}$ where $\mid \mathcal{K} \mid= K$ is the number of users
. The network spectrum between Alice to users is $B$ Hz. 
Our channels model assumption between Alice and legal or illegal receiver such as users and Willie that is represented by set $m \in \{k,w\}$, are based on the path loss and Rayleigh fading model, i.e., $h_{am}=\chi_{am} d_{am}^{-\alpha}$ where $\chi_{am}$ is the channel fading coefficient, $d_{am} >0$ is the 
distance between Alice and $m$, and $\alpha$ is the path loss exponent.
Let $\mathbf{p}=[p_1,\dots,p_K]^T$ be Alice's transmission power to send packets to users. The achievable data rate of each user is given by:
\begin{align}\label{nomarate}
R_{k}=\log(1+\frac{h_{ak}p_{k}}{h_{ak}\sum_{j=k+1}^{K}p_{j}+\sigma^{2}}),
\end{align}
where $\sigma^{2}$ is the AWGN at all legal receivers, which is assumed to be the same for all users, and the interference term $h_{ab_k}\sum_{j=k+1}^{K}p_{j}$ is due to the PD-NOMA transmission technology. 
\subsection{Hypothesis Testing at Willie}
In the covert communication scenario, Willie as an eavesdropper has to decide whether Alice transmitted the packet to users or not. It is a common assumption that Willie has complete knowledge of Alice's location and this is the worst case of covert communication \cite{21}. 
Willie takes a decision based on the received signal $y_w$, as follows:
\begin{equation}\label{recivedpacketwillie}
y_w= 
\begin{cases}
n_w, &\quad \mathcal{H}_0,\\
h_{aw}\sqrt{p_a}s_k+n_w, &\quad  \mathcal{H}_1,\\ 
\end{cases}
\end{equation}
where $\mathcal{H}_0$ and $\mathcal{H}_1$ are denoted as the null and the alternative hypothesis to state whether Alice transmitted information $s_k$ with power transmission $p_a$ or not, respectively, and $n_w$ is the AWGN at Willie with variance $\sigma_w^2$.
Thus, the power of each received packet $T_w$ at Willie, is :
\begin{align}
T_w=|y_w|^2 \mathop\gtrless_{\psi_0}^{\psi_1} \vartheta,
\end{align}
where $\vartheta$ is the detection threshold, and $\psi_0$ and $\psi_1$ are the decisions in favor of $\mathcal{H}_0$ and $\mathcal{H}_1$, respectively. 
In dynamic environments, transceivers may not have accurate knowledge of AWGN power. We consider uncertainty for Willie's noise with a known uniform distribution, that $\sigma_{w,db}^2 \in [\check{\sigma}_{db}^2-\mu_{db},\check{\sigma}_{db}^2+\mu_{db}]$, is the changing value range in the dB domain, and denotes as $\check{\sigma}^2$ nominal noise power, and $\mu$  is the noise uncertainty parameter, that $\mu \geqslant 1$, and $\sigma_{w}^2$ has the following distribution:
\begin{equation}\label{sigmaWillie}
f_{\sigma_{w}^2}(x)=\begin{cases}
\frac{1}{2\ln(\mu)x}, &\quad \frac{1}{\mu}\check{\sigma}^2 \leqslant x \leqslant \mu\check{\sigma}^2,\\
0, &\quad  \text{o.w}.\\ 
\end{cases}
\end{equation}
Based on problem formulation \eqref{recivedpacketwillie}, if Alice does not transmit a packet, $\mathcal{H}_0$ occurs, whereas Willie decides $\psi_1$ happens, it means false alarm, and the following probability is defined:
\begin{align}\label{falsalarm}
P_{FA} \triangleq \text{Pr}\{\psi_1 | \mathcal{H}_0 \} =\text{Pr}\{\sigma_w^2 \geqslant \vartheta\}.
\end{align}
Similarly, based on \eqref{recivedpacketwillie}, if Alice transmits a packet to users, $\mathcal{H}_1$ occurs, whereas Willie decides $\psi_0$ happens, it means he missed the detection, and the following probability is defined:
\begin{align}\label{missdetection}
P_{MD} \triangleq \text{Pr}\{\psi_0 | \mathcal{H}_1 \} = \text{Pr}\{ p_ah_{aw}
+\sigma_w^2 \leqslant \vartheta\}.
\end{align}
Willie uses radiometer to minimize the total error rate of detection \cite{22}, that is indicated by $\xi$:
\begin{align}
\xi=P_{FA}+P_{MD}.
\end{align}	

\subsection{Detection Performance at Willie}
Willie always hopes to minimize the total error rate $\xi$. Hence, we derive the optimal detection threshold $\vartheta^{\ast}$ from the view point of Willie in order to find the minimum optimal total error rate $\xi^{\ast}$. 
Based on \eqref{sigmaWillie}, \eqref{falsalarm}, and \eqref{missdetection}, the false alarm probability and the miss detection probability at Willie are given by:
\begin{equation}\label{PFA}
P_{FA}=\begin{cases}
1, &\quad  \vartheta < \frac{\check{\sigma}^2}{\mu},\\
\kappa_1, &\quad  \frac{\check{\sigma}^2}{\mu} \leqslant \vartheta \leqslant \check{\sigma}^2 \mu,\\ 
0, &\quad  \vartheta>\check{\sigma}^2 \mu,\\ 
\end{cases}
\end{equation}
\begin{equation}\label{PMD}
P_{MD}=\begin{cases}
0, &\quad  \vartheta <  p_ah_{aw}+\frac{\check{\sigma}^2}{\mu},\\
\kappa_2, &\quad p_ah_{aw}+ \frac{\check{\sigma}^2}{\mu} \leqslant \vartheta \leqslant  p_ah_{aw}+\check{\sigma}^2 \mu,\\ 
1, &\quad  \vartheta> p_ah_{aw}+\check{\sigma}^2 \mu,\\  
\end{cases}
\end{equation}
where
$\kappa_1=\int_{\vartheta}^{\check{\sigma}^2 \mu}\frac{1}{2\ln(\mu)x}dx=\frac{1}{2\ln(\mu)}\ln(\frac{\check{\sigma}^2 \mu}{\vartheta}),$
and
$\kappa_2=\int_{\frac{\check{\sigma}^2}{\mu}}^{\vartheta- p_ah_{aw}}\frac{1}{2\ln(\mu)x}dx=\frac{\ln(\frac{\mu(\vartheta- p_ah_{aw})}{\check{\sigma}^2})}{2\ln(\mu)}.$

$\mathbf{Theorem 1:}$The optimal detection threshold $\vartheta^{\ast}$, which can minimize $\xi$ at Willie, is equal to $ p_ah_{aw}+\frac{\check{\sigma}^2}{\mu}$, and the corresponding minimum total error rate is given by 
$\xi^{\ast}=\frac{1}{2\ln(\mu)}\ln(\frac{\mu\sigma_w^2}{p_ah_{aw}+\frac{\sigma_w^2}{\mu}})$.

$Proof:$ The detailed proof is reported in Appendix. 
$\blacksquare$

According to the minimum achievable total error rate from the perspective of Willie, we design the transmit power of Alice $p_a=\sum_{k=1}^{K}p_{k}$, to ensure the minimum total error rate $\xi^{\ast}$ at Willie to being no less than the specific value, i.e, $\xi^{\ast} \geqslant 1-\epsilon_w$, where $\epsilon_w$ is an arbitrarily small value.
\subsection{AoI of Received Packets at users}
In the proposed system model as mentioned above, Alice periodically generates packets and covertly transmit them to users at the same time. 
Let $\mathbf{d}=[d_1,\dots,d_K]^T$ be the age of freshness of packets that is requested by users. We use AoI as a measure of packets' freshness that is requested by the $k$-th user at time $t$, and is defined as:
\begin{align}\label{ageformula}
d_k(t)= t-a_k(t),
\end{align}
where $a_k(t)$ is the generation time of the newest requested packet by $k$-th user, and we map $d_k(t)$ to $d_k$ \cite{1}. We applied the $just-in-time$ transmission policy to distinguish the minimum achievable AoI \cite{1}. In this policy, Alice quickly generates a new update packet and starts its service time just after user receives the current update packet in service. Then, the average AoI that users receive the packet is:
\begin{align}\label{age}
\bar{A}_a= \frac{1}{K}\sum_{k=1}^{K}d_k.
\end{align}
\subsection{Guarantee Reliable Covert Communication with Desirable AoI}
Guaranteeing a reliable covert communication with desirable minimum AoI in the time-varying channel, is not always possible under the above-mentioned constraints. \textcolor{black}{As long as Alice is aware of the channel conditions and how much is valid, she will be able to ensure reliable covert transmission.} The validity of knowledge is directly proportional to the AoC, therefore she should send the packet to all users prior to channel variation. Therefore, Alice investigates channel conditions every $\tau$ second, which is present the coherence time and equal to AoC, and the CSI of a wireless link is constant within each $\tau$ \cite{2}.
We propose for the first time a system model that warranties reliable covert communication with desirable minimum AoI by satisfying the following constraint:
\begin{align}\label{conscoherence}
d_{k} \leqslant \tau -\delta,~\forall k \in \mathcal{K},
\end{align}
where $\delta$ is the required time for measuring channel variation, and $\delta<<\tau$.
\begin{figure}[t]
	\centering
	\includegraphics[width=3.5in]{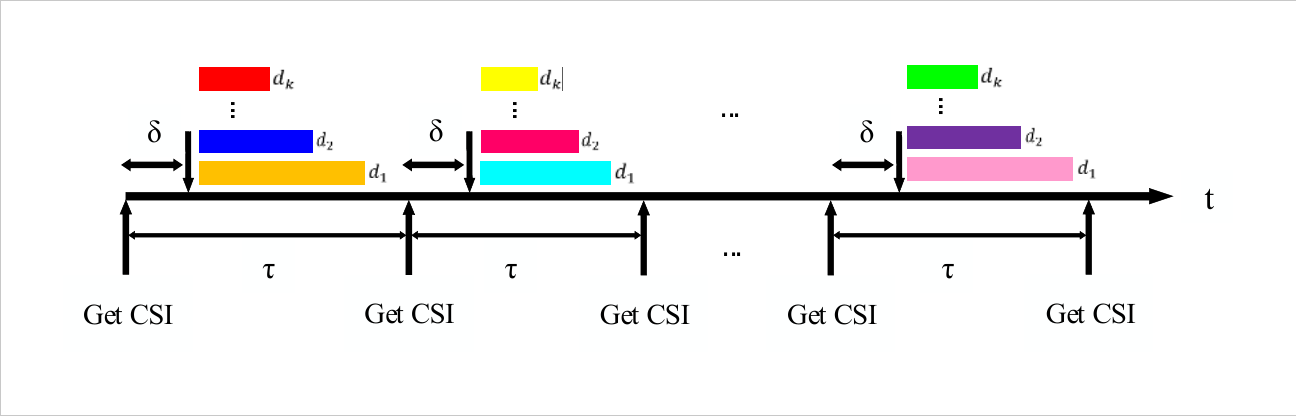}
	\caption{Time framing for proposed constraint with  PD-NOMA assistance.}
	\label{framing}
\end{figure}
\textcolor{black}{Fig. \ref{framing} shows the time framing diagram for the proposed new constraint \eqref{conscoherence}. The PD-NOMA technique makes use of SIC ordering, and rate sorting based on problem formulation \eqref{nomarate}, if $h_{ai} < h_{aj}$, where $\{i,j\} \subset \mathcal{K}$, then $R_i < R_j$, and consequently $d_j < d_i$.}
The result of this constraint causes: 1) Alice can send the requested packets with \textcolor{black}{desirable packet size}. Otherwise, Alice is able to send short packets \cite{14}.
2) \textcolor{black}{Alice is always aware of channel variations, accordingly transmits the requested packet, and redecides the level of power allocation if the related AoI takes more than AoC. Therefore, always guarantees reliable covert communication.}
\section{Problem Formulation and Solution Methodology}
\subsection{Problem Formulation}
To obtain the reliable covert PD-NOMA communication system with desirable AoI, we aim to minimize the average AoI by jointly optimizing the transmit power of Alice and the age of the $k$-th user's packet, subject to the reliable covert constraint, the total transmit power constraint at Alice, the covertness constraint at Willie, the QoS constraint at users. Therefore, the optimization problem is formulated as:
\begin{subequations}\label{obj}
	\begin{align}
	&\min_{\mathbf{p},\mathbf{d}} \bar{A}_a \label{averageageobj}
	\\
	\text{s.t.}~~&d_{k} \leqslant \tau -\delta,~\forall k \in \mathcal{K},\label{coherence}
	\\
	&\sum_{k=1}^{K}p_k\leqslant P_{\text{max}},\label{power}
	\\
	&\xi^{\ast} \geqslant 1-\epsilon_w,\label{covert}
	\\
	&d_{k}\log_2(1+\frac{h_{ak}p_{k}}{h_{ak}\sum_{j=k+1}^{K}p_{j}+\sigma^{2}}) \geqslant \frac{S}{B},~\forall k \in \mathcal{K},\label{rate}
	\end{align}
\end{subequations}
where \eqref{coherence} guarantees the reliable covert communication subject to the AoI of each user to be less than AoC, \eqref{power} is confirms that the transmit power of Alice to its serving all users does not exceed the power transmission budget $P_{\text{max}}$, and constraint \eqref{covert} guarantees covert transmission from Alice to all users. According to Shannon's formula, the maximum achievable rate at users by guaranteeing the successful delivery of packets from Alice to user is expressed as \eqref{rate}, where $S$ is the size of each update packet. The optimization problem \eqref{obj}, due to the non-convex constraint \eqref{covert} and \eqref{rate}, is challenging to solve. To tackle this challenge, in the following, we optimize the transmit power of Alice and AoI of packets alternatively, until \textcolor{black}{convergence is achieved}.

\subsection{Solution Methodology}
Towards making more tractable problem \eqref{obj}, we propose an efficient iterative algorithm. We optimize $d_k$ for given $p_a$ by solving a linear programming (LP), and $p_a$ for any given $d_k$ based on the successive convex approximation (SCA) optimization technique.
\subsubsection{AoI Optimization}
To find the AoI for any given $p_a$, \eqref{obj} can be solved by optimizing the following problem:
\begin{subequations}\label{subobj1}
	\begin{align}
	&\min_{\mathbf{d}} \bar{A}_a \label{subaverageageobj1}
	\\
	\text{s.t.}~~&d_{k} \leqslant \tau -\delta,~\forall k \in \mathcal{K},\label{subcoherence1}
	\\
	&d_{k}\log_2(1+\frac{h_{ak}p_{k}}{h_{ak}\sum_{j=k+1}^{K}p_{j}+\sigma^{2}}) \geqslant \frac{S}{B},~\forall k \in \mathcal{K}.\label{subrate1}
	\end{align}
\end{subequations}
Since the problem \eqref{subobj1} has standard LP form, so it can be solved by existing optimization tools such as CVX \cite{20}.
\subsubsection{Reliable Covert Communication Optimization}
To find the optimal transmit power Alice $p_a$ for given $d_k$, \eqref{obj} can be optimized by solving the following problem:
\begin{subequations}\label{subobj2}
	\begin{align}
	&\min_{\mathbf{p}} \bar{A}_a \label{subaverageageobj2}
	\\
	\text{s.t.}~~&\sum_{k=1}^{K}p_k\leqslant P_{\text{max}},\label{subpower2}
	\\
	&\frac{1}{2\ln(\mu)}\ln(\frac{\mu\sigma_w^2}{p_ah_{aw}+\frac{\sigma_w^2}{\mu}})\geqslant 1-\epsilon_w,\label{subcovert2}
	\\
	&d_{k}\log_2(1+\frac{h_{ak}p_{k}}{h_{ak}\sum_{j=k+1}^{K}p_{j}+\sigma^{2}}) \geqslant \frac{S}{B},~\forall k \in \mathcal{K}.\label{subrate2}
	\end{align}
\end{subequations}
We write constraint \eqref{covert} as \eqref{subcovert2} following the Appendix in problem \eqref{subobj2}. However, should note that it is not easy to solve problem \eqref{subobj2} because of constraint \eqref{subrate2} that is non-convex. In the following, with respect to the first-order Taylor expansion, we rewrite the term $R_{k}$ in constraint \eqref{subrate2} as follows:
\begin{align}\label{SCAaproximation}
R_{k}&=\log_2(1+\frac{h_{ak}p_{k}}{h_{ak}\sum_{j=k+1}^{K}p_{j}+\sigma^{2}})\nonumber \\
&=\log_2(h_{ak}\sum_{j=k}^{K}p_{j}+\sigma^{2})-\log_2(h_{ak}\sum_{j=k+1}^{K}p_{j}+\sigma^{2}), \nonumber \\
&\geqslant\log_2(h_{ak}\sum_{j=k}^{K}p_{j}+\sigma^{2})-\log_2(h_{ak}\sum_{j=k+1}^{K}p_{j}^{\iota}+\sigma^{2})\nonumber \\
&-\sum_{j=k+1}^K\frac{h_{ak}\log_2(e)}{h_{ak}\sum_{j=k+1}^{K}p_{j}^{\iota}+\sigma^{2}}[p_j-p_j^{\iota}],
\end{align}
where $p_{j}^{\iota}$ is the power of the $j$-th user in the $\iota$-th iteration.
According to problem \eqref{SCAaproximation}, constraint \eqref{subrate2} is rewritten as follows:
\begin{align}\label{subrate2SCAaproximation}
d_{k}(\log_2(h_{ak}\sum_{j=k}^{K}p_{j}+\sigma^{2})-\log_2(h_{ak}\sum_{j=k+1}^{K}p_{j}^{\iota}+\sigma^{2})\nonumber \\
-\sum_{j=k+1}^K\frac{h_{ak}\log_2(e)}{h_{ak}\sum_{j=k+1}^{K}p_{j}^{\iota}+\sigma^{2}}[p_j-p_j^{\iota}]) \geqslant \frac{S}{B},~\forall k \in \mathcal{K}.
\end{align}
Problem \eqref{subobj2} is approximated as the following problem:
\begin{subequations}\label{subobj21}
	\begin{align}
	&\min_{\mathbf{p}} \bar{A}_a \label{subaverageageobj21}
	\\
	\text{s.t.}~~&,\eqref{subpower2}, \eqref{subcovert2}, \eqref{subrate2SCAaproximation}\label{subpower21}.
	\end{align}
\end{subequations} 
Based on the proposed solution methodology, the original problem \eqref{obj} is partitioned into two optimization sub-problems \eqref{subobj1} and \eqref{subobj21} that are solved iteratively. Furthermore, the obtained solution in each iteration is used as the input of the next iteration, and can be efficiently solved by software toolboxes like CVX \cite{20}.

\section{Simulation and Results}
In this section, we present numerical results to evaluate the performance of the proposed reliable covert PD-NOMA communication with minimum AoI system. We assume users, and Willie are randomly and uniformly distributed in the coverage area with a radius of $R=100$ m, and Alice located at the center of area. The network parameters are $\tau=10$ ms, $B=1$ MHz, $S=1$ Kbits, $\epsilon_w=0.95$, $\sigma_w^2=-120$ dB, $\sigma^2=-160$ dB, $\mu=3$ dB, $\alpha \in [2.5-3.5]$.

\begin{figure}[h]
	\centering
	\includegraphics[width=3.5in]{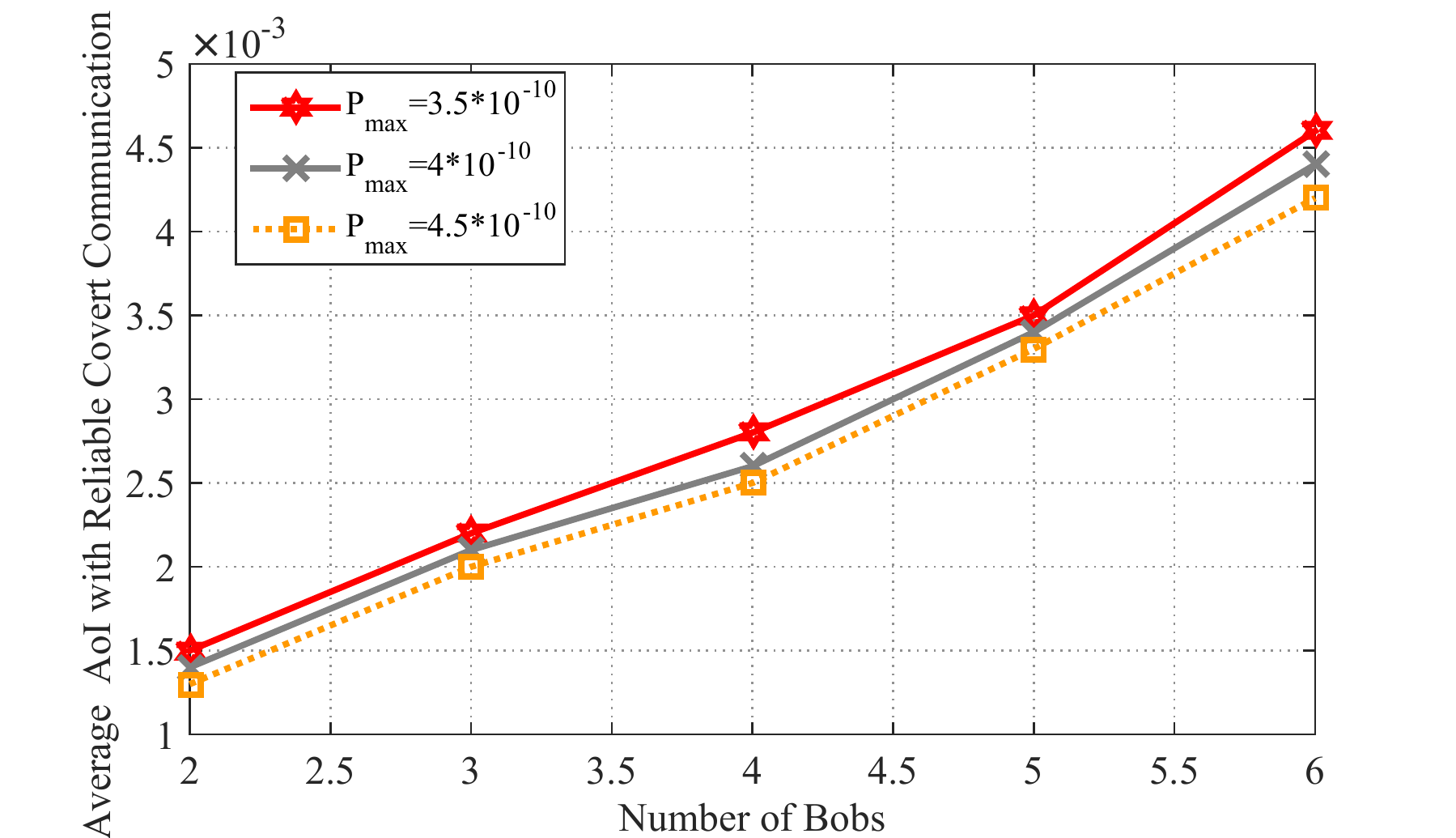}
	\caption{Average reliable covert AoI versus different number of users.}
	\label{avgcagenumBobs}
\end{figure}

In Fig.\ref{avgcagenumBobs}, we demonstrate the average AoI with reliable covert communication versus the number of users for different amounts of power transmission budget, $P_{\text{max}}$. In the optimization problem \eqref{obj}, constraint \eqref{rate} is limited by the fixed lower band $S/B$. \textcolor{black}{
It means, by increasing the number of users i.e., $K=\{2,3,4,5,6\}$, the achievable data rate of each user is decreasing, and accordingly the AoI of each user $d_k$, and the average AoI, $\bar{A}_a$ are increasing, which is confirmed in \eqref{rate}. Therefore, by increasing the number of users in the network, AoI increases.} In this figure, we observe that with different amounts of $P_{\text{max}}$, by increasing the number of users, the AoI are increasing. \textcolor{black}{Increased transmitter power budget allocates more power to users, resulting in higher data rates and lower AoI. Hence, more power budget allocation in the network leads to decreased AoI.}

\begin{figure}[h]
	\centering
	\includegraphics[width=3.5in]{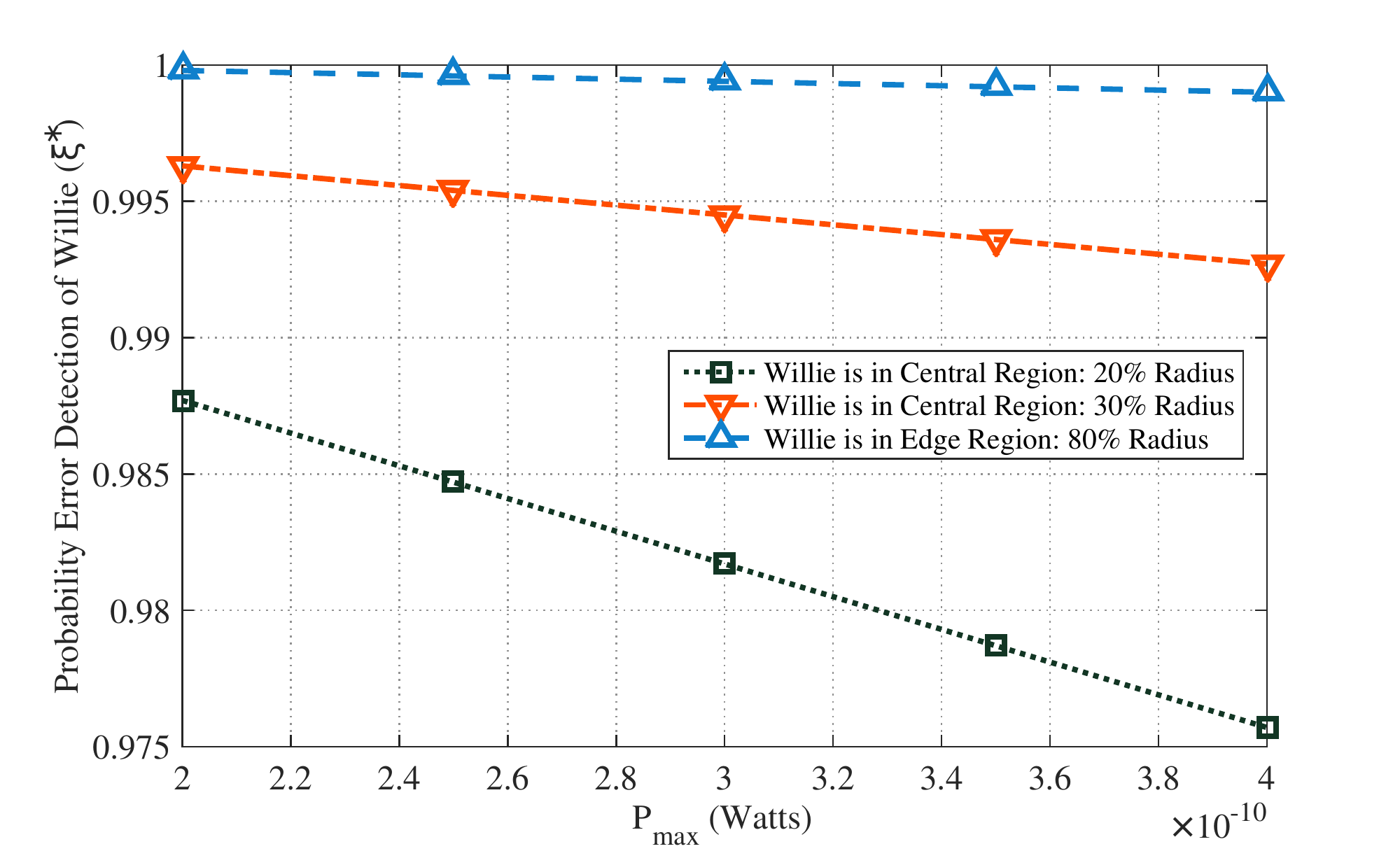}
	\caption{Probability of error detection of Willie $\xi^{\star}$ versus variation of $P_{\text{max}}$ for $K=3$.}\label{errorprobpoweralic}
\end{figure}

In Fig. \ref{errorprobpoweralic}, we show the total error rate, $\xi^{\star}$, versus power transmission budget, $P_{\text{max}}$, with different locations of Willie. \textcolor{black}{$\xi^{\star}$ is monotonically decreasing with increasing $P_{\text{max}}$. Thus, by increasing $P_{\text{max}}$, Willie can detect users’s covert communication easily.} It is observed by increasing $P_{\text{max}}$, $\xi^{\star}$ goes toward 0 gradually, which is consistent with our findings in Theorem 1. In addition, the result in Fig. \ref{errorprobpoweralic} is matched with the problem formulation \eqref{subobj2}, constraint \eqref{subcovert2}, thus verifying the accuracy of the derived analytical results. On the other hand, the distance between Willie and Alice is contrary to the total error rate $\xi^{\star}$. \textcolor{black}{As a result, the probability of error detection decreases as the Willie moves away from the central region to the edge region.}

\begin{figure*}[!t]
	\centering
	\subfloat[]{\includegraphics[width=3.5in]{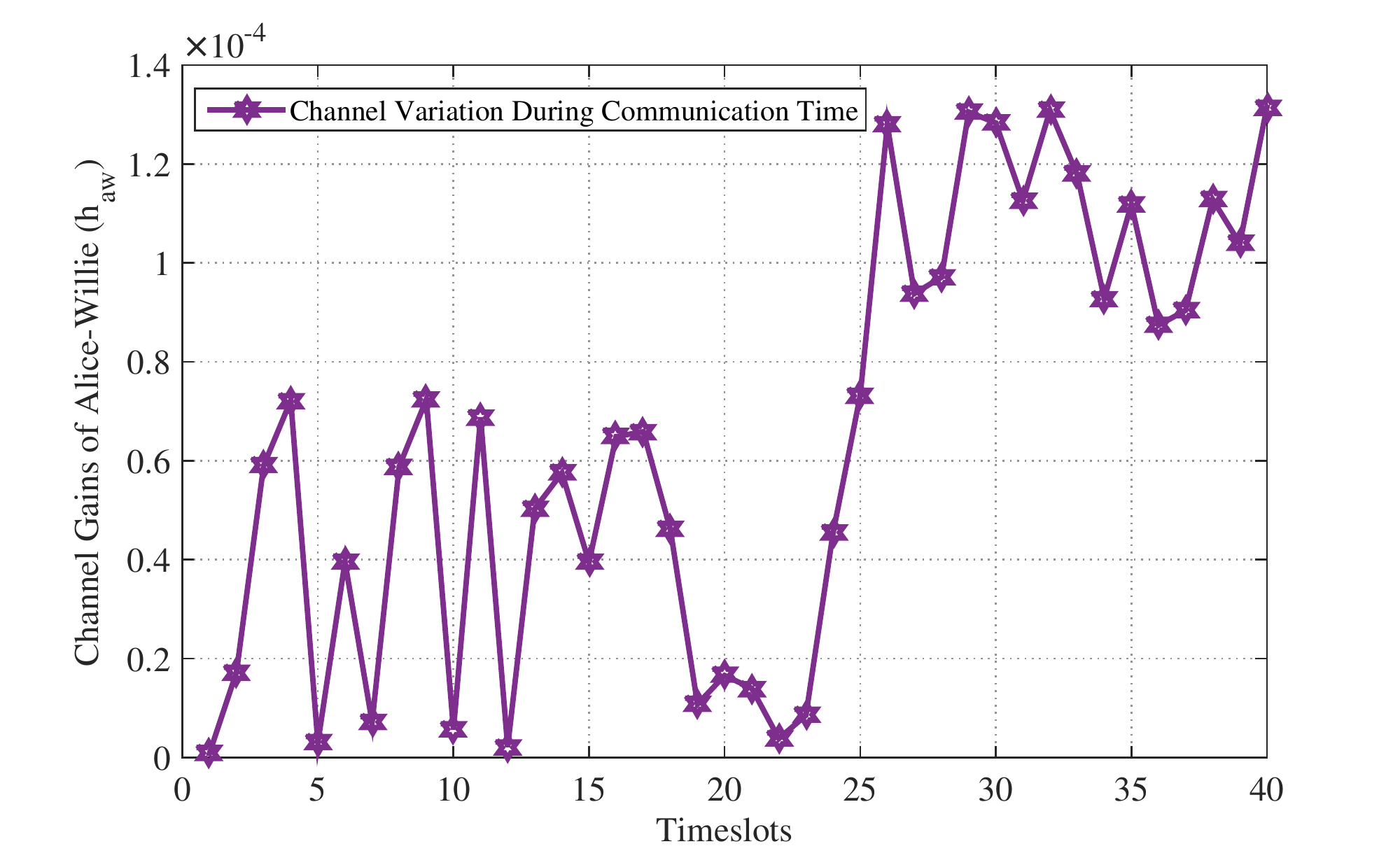}%
		\label{channelgainalicewiilieperslot}}
	\hfil
	\subfloat[]{\includegraphics[width=3.5in]{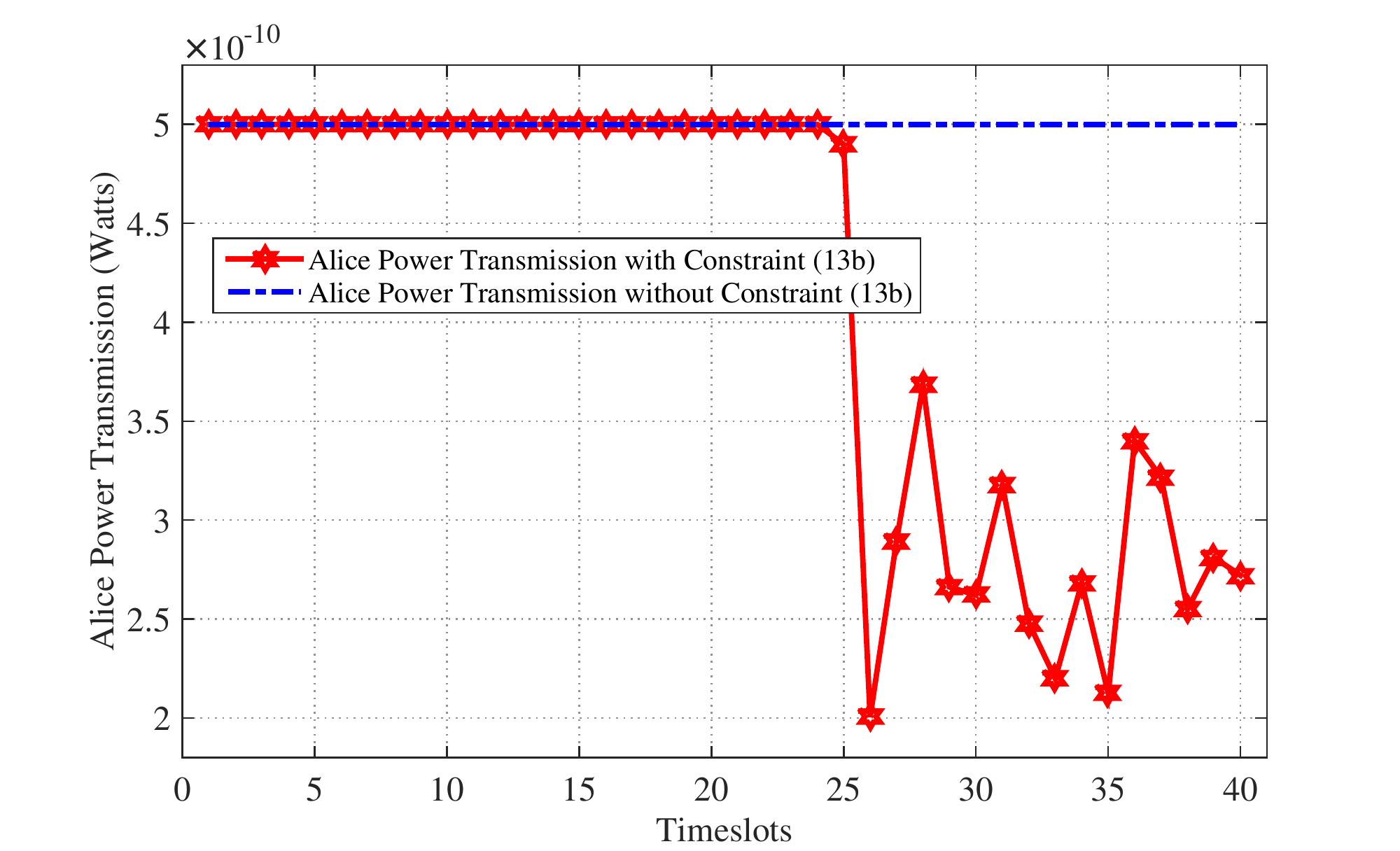}%
		\label{powerperslot}}
	\hfil
	\subfloat[]{\includegraphics[width=3.5in]{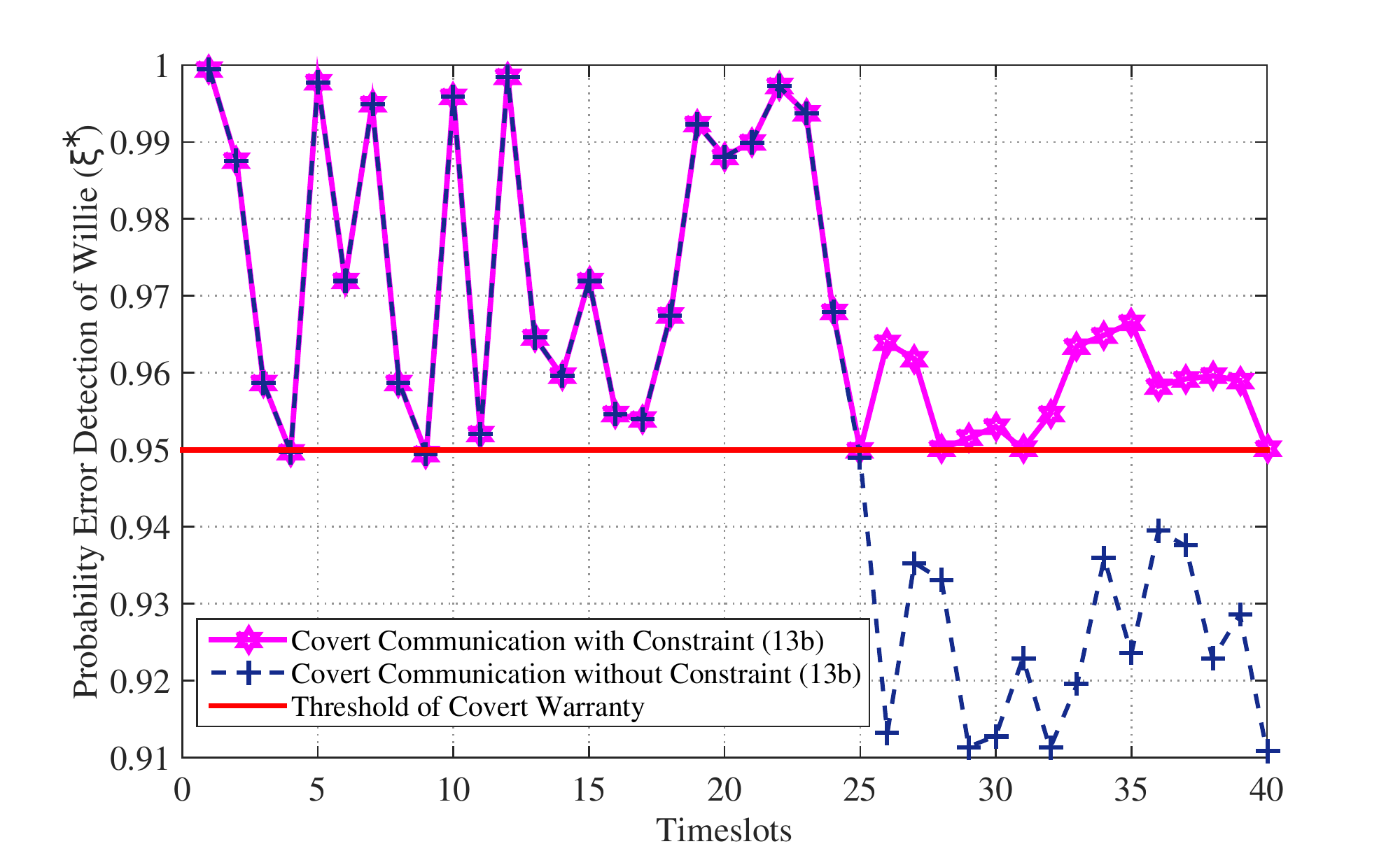}%
		\label{covertcovertness}}
	\caption{The presentation of channel variation during different AoC, and Alice responsibility to set the power transmission proper to the channel variation, and show the performance of proposed new constraint \eqref{coherence} for $K=3$. (a) Channel gains of Alice to Willie, (b) Alice power transmission $P_a$, (c) comparison between reliable/ unreliable covert communication with/without constraint \eqref{coherence}, vs different time slots.}
	\label{final}
\end{figure*}

In Fig. \ref{final}, we represent the necessity of constraint \eqref{coherence}. As mentioned above, this constraint guarantees reliable covert communication with desirable minimum AoI.
In the absence of \eqref{coherence}, Alice has a blind channel knowledge that
covertness precision may be finite or non-exist. We solve our problem formulation \eqref{subobj21}, and the average AoI is $d_k=0.4$ s, which is more than AoC. Therefore, Alice should be divided the packet into some nats \cite{14}, and send them in iterative time slots $n$, such $n=d_k/\tau$. As a result, Alice monitors the channel variation which is caused by channel fading model or Willie's movement between different time slots (but be fixed in duration time slot) and decides based on guaranteeing reliable covert communication. \textcolor{black}{Fig. \ref{channelgainalicewiilieperslot}, shows the channel variation in communication time, although Alice is unaware if constraint \eqref{coherence} is ignored.}
According to channel variation observation, Alice allocates power in the way of preserving the required total error rate of Willie's detection to users. In Fig. \ref{powerperslot}, we show the Alice power transmission decision based on channel variation in different time slots. In the case of considering constraint \eqref{coherence}, Alice power allocation is appropriate to channel variation with the aim of reliable covert rate protection, while without considering constraint \eqref{coherence}, Alice allocates power with steady state. In Fig. \ref{covertcovertness}, we clarify the impact of constraint \eqref{coherence} on the reliable covert wit minimum AoI. In the covert communication, the channel gain between Alice and Willie, and the power level of Alice are two effective parameters for Willie's detection. With the existence of constraint \eqref{coherence}, Alice monitors the channel variation and sets the power allocation to users properly to satisfy QoS, minimum AoI, and reliable covert communication, jointly. The blue dash line acknowledges that without considering constraint \eqref{coherence}, in some time slots Willie crosses from the threshold line of covert warranty. Moreover, it can detect that there is a communication between Alice and users, hence Alice guarantees the covert communication incorrectly. 

\section{Conclusion}
In this letter, we proposed a reliable covert communication with desirable minimum AoI in time varying channels. We apply PD-NOMA as a transmission technology to improve the covertness and AoI by adopting superposition coding at the transmitter and SIC ordering at the receiver, correspondingly. 
A new constraint represents the requirement of jointly reliable covert communication and minimum AoI, which expresses AoI should be less than AoC. In other words, Alice is responsible for answering the requested packet based on their QoSs, and the maximum power transmission budgets with reliable covert communication in a minimum AoI manner, While Alice should find the AoC, which is equivalent to the coherence time, so divide the AoI of each packet $d_k$, into the AoC $\tau$, and send iteratively in each time slot $n$. Hence, Alice monitors the channel condition at the start of each time slot and jointly transmits users's requests in a reliable covert and fresh manner.

\appendices
\section{Proof of Theorem 1}\label{covertappendix}
As shown in \eqref{PFA} and \eqref{PMD}, the total error rate $\xi$ is given by:
\begin{equation}\label{xi}
\xi=\begin{cases}
1, &\quad  \vartheta< \frac{\check{\sigma}^2}{\mu},\\
\kappa_1, &\quad  \frac{\check{\sigma}^2}{\mu} \leqslant \vartheta < p_ah_{aw}+\frac{\check{\sigma}^2}{\mu},\\
\kappa_1+\kappa_2, &\quad  p_ah_{aw}+\frac{\check{\sigma}^2}{\mu} \leqslant \vartheta \leqslant \check{\sigma}^2 \mu\\
\kappa_2, &\quad \check{\sigma}^2 \mu < \vartheta \leqslant  p_ah_{aw}+\check{\sigma}^2 \mu,\\ 
1, &\quad  \vartheta> p_ah_{aw}+\check{\sigma}^2 \mu.\\  
\end{cases}
\end{equation}
It is obvious that $\xi$ in case of $\frac{\check{\sigma}^2}{\mu} \leqslant \vartheta < p_ah_{aw}+\frac{\check{\sigma}^2}{\mu}$ monotonically decreasing function of $\vartheta$. As such, the optimal detection threshold is obtained by $\vartheta^{\ast}= p_ah_{aw}+\frac{\check{\sigma}^2}{\mu}$ and the corresponding minimum total error rate is  $\xi^{\ast}=\frac{1}{2\ln(\mu)}\ln(\frac{\mu\sigma_w^2}{p_ah_{aw}+\frac{\sigma_w^2}{\mu}})$.

\bibliography{citation_AoI-Cov}	

\begin{thebibliography}{10}
\providecommand{\url}[1]{#1}
\csname url@samestyle\endcsname
\providecommand{\newblock}{\relax}
\providecommand{\bibinfo}[2]{#2}
\providecommand{\BIBentrySTDinterwordspacing}{\spaceskip=0pt\relax}
\providecommand{\BIBentryALTinterwordstretchfactor}{4}
\providecommand{\BIBentryALTinterwordspacing}{\spaceskip=\fontdimen2\font plus
\BIBentryALTinterwordstretchfactor\fontdimen3\font minus
  \fontdimen4\font\relax}
\providecommand{\BIBforeignlanguage}[2]{{%
\expandafter\ifx\csname l@#1\endcsname\relax
\typeout{** WARNING: IEEEtran.bst: No hyphenation pattern has been}%
\typeout{** loaded for the language `#1'. Using the pattern for}%
\typeout{** the default language instead.}%
\else
\language=\csname l@#1\endcsname
\fi
#2}}
\providecommand{\BIBdecl}{\relax}
\BIBdecl

\bibitem{6}
V.-L. Nguyen, P.-C. Lin, B.-C. Cheng, R.-H. Hwang, and Y.-D. Lin, ``Security
  and privacy for {6G}: A survey on prospective technologies and challenges,''
  \emph{IEEE Communications Surveys Tutorials}, vol.~23, no.~4, pp. 2384--2428,
  2021.

\bibitem{5}
X.~Jiang, X.~Chen, J.~Tang, N.~Zhao, X.~Y. Zhang, D.~Niyato, and K.-K. Wong,
  ``Covert communication in {UAV}-assisted air-ground networks,'' \emph{IEEE
  Wireless Communications}, vol.~28, no.~4, pp. 190--197, 2021.

\bibitem{7}
T.-T. Chan, H.~Pan, and J.~Liang, ``Age of information with joint packet coding
  in industrial {IoT},'' \emph{IEEE Wireless Communications Letters}, vol.~10,
  no.~11, pp. 2499--2503, 2021.

\bibitem{8}
R.~D. Yates, Y.~Sun, D.~R. Brown, S.~K. Kaul, E.~Modiano, and S.~Ulukus, ``Age
  of information: An introduction and survey,'' \emph{IEEE Journal on Selected
  Areas in Communications}, vol.~39, no.~5, pp. 1183--1210, 2021.

\bibitem{9}
M.~Samir, M.~Elhattab, C.~Assi, S.~Sharafeddine, and A.~Ghrayeb, ``Optimizing
  age of information through aerial reconfigurable intelligent surfaces: A deep
  reinforcement learning approach,'' \emph{IEEE Transactions on Vehicular
  Technology}, vol.~70, no.~4, pp. 3978--3983, 2021.

\bibitem{12}
R.~D. Yates, Y.~Sun, D.~R. Brown, S.~K. Kaul, E.~Modiano, and S.~Ulukus, ``Age
  of information: An introduction and survey,'' \emph{IEEE Journal on Selected
  Areas in Communications}, vol.~39, no.~5, pp. 1183--1210, 2021.

\bibitem{14}
W.~Yang, X.~Lu, S.~Yan, F.~Shu, and Z.~Li, ``Age of information for
  short-packet covert communication,'' \emph{IEEE Wireless Communications
  Letters}, vol.~10, no.~9, pp. 1890--1894, 2021.

\bibitem{15}
Y.~Wang, S.~Yan, W.~Yang, and Y.~Cai, ``Covert communications with constrained
  age of information,'' \emph{IEEE Wireless Communications Letters}, vol.~10,
  no.~2, pp. 368--372, 2021.

\bibitem{17}
L.~Lv, Q.~Wu, Z.~Li, Z.~Ding, N.~Al-Dhahir, and J.~Chen, ``Covert communication
  in intelligent reflecting surface-assisted {NOMA} systems: Design, analysis,
  and optimization,'' \emph{IEEE Transactions on Wireless Communications},
  vol.~21, no.~3, pp. 1735--1750, 2022.

\bibitem{16}
L.~Tao, W.~Yang, S.~Yan, D.~Wu, X.~Guan, and D.~Chen, ``Covert communication in
  downlink {NOMA} systems with random transmit power,'' \emph{IEEE Wireless
  Communications Letters}, vol.~9, no.~11, pp. 2000--2004, 2020.

\bibitem{19}
M.~Wang, W.~Yang, X.~Lu, C.~Hu, B.~Liu, and X.~Lv, ``Channel inversion power
  control aided covert communications in uplink noma systems,'' \emph{IEEE
  Wireless Communications Letters}, pp. 1--1, 2022.

\bibitem{10}
Q.~Wang, H.~Chen, C.~Zhao, Y.~Li, P.~Popovski, and B.~Vucetic, ``Optimizing
  information freshness via multiuser scheduling with adaptive {NOMA}/{OMA},''
  \emph{IEEE Transactions on Wireless Communications}, vol.~21, no.~3, pp.
  1766--1778, 2022.

\bibitem{11}
R.~S. B.~A. G, S.~Deshmukh, S.~R.~B. Pillai, and B.~Beferull-Lozano, ``Energy
  efficient {AoI} minimization in opportunistic {NOMA}/{OMA} broadcast wireless
  networks,'' \emph{IEEE Transactions on Green Communications and Networking},
  pp. 1--1, 2021.

\bibitem{18}
------, ``Energy efficient {AoI} minimization in opportunistic {NOMA}/{OMA}
  broadcast wireless networks,'' \emph{IEEE Transactions on Green
  Communications and Networking}, pp. 1--1, 2021.

\bibitem{21}
X.~Jiang, Z.~Yang, N.~Zhao, Y.~Chen, Z.~Ding, and X.~Wang, ``Resource
  allocation and trajectory optimization for {UAV}-enabled multi-user covert
  communications,'' \emph{IEEE Transactions on Vehicular Technology}, vol.~70,
  no.~2, pp. 1989--1994, 2021.

\bibitem{22}
M.~Forouzesh, P.~Azmi, N.~Mokari, and D.~Goeckel, ``Covert communication using
  null space and 3d beamforming: Uncertainty of willie's location
  information,'' \emph{IEEE Transactions on Vehicular Technology}, vol.~69,
  no.~8, pp. 8568--8576, 2020.

\bibitem{1}
M.~A. Abd-Elmagid and H.~S. Dhillon, ``Average peak age-of-information
  minimization in {UAV}-assisted {IoT} networks,'' \emph{IEEE Transactions on
  Vehicular Technology}, vol.~68, no.~2, pp. 2003--2008, 2019.

\bibitem{2}
Q.~Zhang, W.~Saad, and M.~Bennis, ``Millimeter wave communications with an
  intelligent reflector: Performance optimization and distributional
  reinforcement learning,'' \emph{IEEE Transactions on Wireless
  Communications}, pp. 1--1, 2021.

\bibitem{20}
M.~Grant and S.~Boyd, ``{CVX}: {MATLAB} software for disciplined convex
  programming, version 2.1,'' 2014.

\end{thebibliography}
\bibliographystyle{ieeetran}

\ifCLASSOPTIONcaptionsoff
  \newpage
\fi






\end{document}